# Induced ferromagnetic moment at the interface between epitaxial $SrRuO_3$ film and $Sr_2RuO_4$ single crystal

Seung Ran Lee[1,2,*], Muhammad Shahbaz Anwar[3], Yeong Jae Shin[1,2], Min-Cheol Lee[1,2], Yusuke Sugimoto[3], Masanao Kunieda[3], Shingo Yonezawa[3], Yoshiteru Maeno[3] & Tae Won Noh[1,2]

[1] Center for Correlated Electron Systems, Institute for Basic Science (IBS), Seoul 151-747, Republic of Korea

[2] Department of Physics and Astronomy, Seoul National University, Seoul 151-747, Republic of Korea

[3] Department of Physics, Graduate School of Science, Kyoto University, Kyoto 606-8502, Japan

**Abstract**

$SrRuO_3$ (SRO113) is an important material for device physics particularly as one of the best metallic oxide electrodes for ferroelectric devices. This oxide has moderate electron correlations with novel properties including ferromagnetic ordering, which can be utilized in future to spintronics and superconducting spintronics devices. Recently, we observed strongly enhanced magnetization of SRO113 thin films grown on single crystals of the spin-triplet superconductor $Sr_2RuO_4$ (SRO214). To clarify the origin of such an enhancement, we conducted systematic investigations of magnetic properties of SRO113 films deposited on a variety of oxide substrates. We carefully subtracted the substrate contributions and found that the enhanced magnetization occurs only for SRO113/SRO214 films. We further found that neither strain nor metallicity of the substrate plays any significant roles in the enhancement. The X-ray magnetic circular dichroism reveals that the substrate-induced strain does not switch the $Ru^{4+}$ state from the low-spin to high-spin states. The film-thickness dependence of the magnetization of SRO113/SRO214 films strongly suggest that the additional magnetization arises due to the induction of magnetic moment into the SRO214 substrate over 20-nm depth. Our results imply new magnetic functionality that can trigger studies searching for yet unknown physical phenomena in magnetic ruthenates.

## INTRODUCTION

$SrRuO_3$ (SRO113) thin films have been extensively investigated among its family of the $Sr_{n+1}Ru_nO_{3n+1}$ Ruddlesden-Popper (RP) series due to its versatile functionalities for oxide-based device applications and distinctive electrical and magnetic properties compared with other transition-metal oxides. SRO113 has chemically stable surface[1]. Its orthorhombic perovskite structure has especially high affinity to other functional perovskite-type oxides such as ferromagnetic, ferroelectric, and superconducting systems[2-5]. Below its Curie temperature ($T_{Curie}$) ≈160 K, this oxide exhibits itinerant ferromagnetism, which is rare among the 4*d* oxides. However, in strained films, $T_{Curie}$ is known to always decrease[6]. This fact indicates that magnetic properties of SRO113 are sensitive to its structural distortion such as the $RuO_6$ octahedral distortion[1, 7] .

As a good approximation, Ru ions in SRO113, taking the 4+ valence with the $4d^4$ configuration, exhibit the low-spin state in both film and bulk forms due to the large crystal-field splitting between the $t_{2g}$ and $e_g$ levels. Thus, the magnetic moment of a $Ru^{4+}$ ion is naively expected to be $2\mu_B$. Taking into account various interactions such as spin-orbit coupling[8], *d-p* hybridization[9, 10], and electron-electron correlation[11, 12], the magnetization *M* of SRO113 is calculated to become smaller. Thus, $2\mu_B$/Ru has been known as the upper limit of *M*. Many theoretical approaches have also pointed out critical roles of $RuO_6$ octahedral distortions, namely octahedral tilting and rotations[7, 13, 14]. Experimentally, such $RuO_6$ structural changes can be easily induced by strain engineering, i.e. choosing perovskite-oxide single-crystal substrates with different lattice constants, structures, surface terminations, and surface step-size. For example, a perfectly $TiO_2$-terminated $SrTiO_3$ (STO) is one of the most widely used substrates, which yields compressive strain of 0.64%[15]. Highly-crystalline SRO113 films epitaxially grown on STO exhibits *M* of 1.1-1.6 $\mu_B/Ru^{4+}$ [16] and $T_{Curie}$ ~ 150 K[17]. On the other hand, when SRO113 experiences a large tensile strain as in the case of SRO113 on $KTaO_3$, *M* decreases and the easy axis changes from the out-of-plane to the in-plane direction[18]. Note that we also cannot rule out possibility that the quality of films in each report, such as crystallinity and stoichiometry, may also have affected magnetic properties.

We recently introduced a special substrate for growing SRO113 films: a spin-triplet superconductor $Sr_2RuO_4$ (SRO214), which is not commercially available nor an insulating perovskite. This oxide is the $n = 1$ member of the $Sr_{n+1}Ru_nO_{3n+1}$ RP series, to which SRO113 also belongs. SRO214 [19-21] is known as a paramagnetic metal[20] above the superconducting critical temperature $T_c$ = 1.5 K. We successfully grew epitaxial SRO113 thin films on the cleaved *ab* surface of single crystals of SRO214, even under a relatively large lattice mismatch of 1.5% compressive to the film[22]. In this system, penetration of spin-triplet superconductivity from SRO214 into the ferromagnetic SRO113 has been recently observed and this phenomenon can be utilized for future superconducting spintronics devices[22]. On the other hand, surprisingly, we have revealed that SRO113/SRO214 exhibits incredibly enhanced ferromagnetic magnetization with $M \geq 3$ $\mu_B/Ru^{4+}$ [23]. This observation appears to significantly violate the magnetic upper limit of 2 $\mu_B/Ru^{4+}$. On the other hand, $T_{Curie}$ is observed to be ≈ 160 K, the same value as in the bulk. Considering the known strong dependence of $T_{Curie}$ on strain as well as the large compressive strain experienced by the film on SRO214, the observed full recovery of $T_{Curie}$ to the bulk value is rather unexpected.

In this study, we conducted various experiments to investigate the origin of such unique magnetic character of the SRO113/SRO214 system. In order to prepare SRO113 films with differently distorted $RuO_6$, we utilized various commercially available substrates with different lattice constants, namely, $NdGaO_3$ (NGO), STO, $DyScO_3$ (DSO), and $GdScO_3$ (GSO) (listed in the order of decreasing lattice constants)[24]. With these different substrates, effects of $RuO_6$ octahedral distortion can be studied in a systematic way, by comparing structural properties of our SRO113 films to those of theoretical predictions[7]. However, magnetic measurements reveal that strain effect is less relevant to the enhanced magnetization in SRO113 on SRO214 substrate. We also show that SRO113 inclusion in SRO214 substrates or high-spin states caused by charge transfer or orbital reconstruction cannot explain the observed magnetism enhancement. We suggest a plausible model that magnetic moment induced in the SRO214 side of the SRO113/SRO214 interface plays a crucial role.

## MATERIALS AND METHODS

### Deposition of thin films

SRO113 films were grown by a pulsed laser deposition (PLD) using an excimer laser (Coherent, LPXPro 210F) KrF with the wavelength of 248 nm on various substrates: NGO, SRO214, STO, DSO, and GSO. For a pseudo-cubic substrates NGO, DSO, and GSO, all SRO113 films were grown along the pseudo-cubic (001) directions in a similar way to the (001) STO-substrate case[1, 16]. Surfaces of NGO, DSO, and GSO were prepared by a simple in-house annealing process above 1000°C, resulting in mixed termination on the NGO, DSO, and GSO surfaces. The $TiO_2$-terminated STO surface was prepared with etching in buffered oxide etchant and then an annealing process at high temperatures. The surfaces of the SRO214 substrates were prepared by cleaving along the *ab* plane in air. Such cleaved surface exhibits atomic reconstruction due to $RuO_6$ octahedral rotation[25]. The termination surface is in most cases SrO, but RuO termination may also occur depending on the ambient conditions of cleaving[26].

SRO113 films were grown simultaneously as well as separately on various substrates that verify reproducibility of our samples in every case (see Figure S1). Base pressure of a processing vacuum chamber was maintained at around $2\times10^{-9}$ Torr. Before starting the growth, the chamber was filled with 0.1 Torr $O_2$ gas at room temperature (RT). During the SRO113 deposition, substrate temperature was maintained at 700ºC with the same oxygen pressure. The film thickness was *in-situ* monitored by reflection high-energy electron diffraction (S. R. Lee *et al*., unpublished). After film growth, the substrates were cooled down to RT at the same $O_2$ partial pressure.

## Structural characterization

Structural properties of all films were investigated using a high-resolution X-ray diffractometer (Bruker, D8 Discover). Figure 1(a) shows $\theta$ - $2\theta$ wide scans of SRO113 films grown on various substrates. The clear (00*l*) peaks from each substrate and SRO113 confirm the *c*-axis oriented film growth and are consistent with the epitaxial growth of SRO113 films on every substrate. Here, the cubic or pseudo-cubic structural notations are used, except for SRO214, which has a tetragonal structure with $c$ = 1.274 nm. The positions of the SRO113 (00*l*) peaks with fringe satellites clearly shift from the bulk SRO113 positions (dotted lines) according to the in-plane lattice constants of the substrates. The out-of-plane lattice constant of SRO113 decreases when it is deposited on a substrate like GSO with a larger in-plane lattice constant, and vice versa. The narrow rocking-curve peaks in Figure 1(b) with a full width at half maximum of 0.03 – 0.06º assure high quality of our films.

The in-plane strain of SRO113 films are measured by X-ray reciprocal space mapping (XRSM), which are presented in Figures 2(a) – 2(d) and in Figure S2. The SRO113 peaks are aligned on the same vertical lines as the respective substrate peaks, confirming that our films are all fully strained and epitaxially grown on all the substrates.

**Magnetic properties measured with a SQUID magnetometer**

Magnetic properties of SRO113 films are investigated using a SQUID magnetometer (Quantum Design, MPMS-XL). Here, we describe the procedure for the SRO113/SRO214 films in more detail because accurate background subtraction is crucial for this study. Magnetization-versus-temperature (*M*-vs-*T*) curves shown in Figures 3(a) and 3(b) are obtained under zero field after 1-T field cooling from RT to 4 K with fields along the *c* axis and *a* axis of 30-nm-thich SRO113 films, respectively. The measurements under zero field are necessary to avoid the strong paramagnetic signal of the substrates. The relaxation of *M* after switching off the field at 4 K is rather slow, especially for the field *H* along the *c* axis[27]. Note that for SRO113/SRO214 films, we measured *M* before and after the film deposition on the identical substrates and subtract the substrate contributions to obtain the exact magnetization of SRO113 films alone. Likewise, *M* of the SRO113 films grown on other insulating perovskite substrates were obtained by separately measuring *M* of the substrates only and *M* of the SRO113/substrates for proper background subtraction.

**Electronic properties and XMCD measurements**

In order to identify the electronic states, especially of Ru electrons, X-ray absorption spectroscopy (XAS) and X-ray magnetic circular dichroism (XMCD) measurements were performed with circularly polarized X-rays generated from an elliptically polarized undulator at the 2A beamline in the Pohang Light Source. Data were collected at 100 K in total electron yield mode with a near-normal X-ray beam incidence. For the XMCD measurements, we used X-rays with fixed helicity and alternate the direction of the external magnetic fields ($\mu_0 H$ = 0.8 T) to ensure a constant beam profile. We cooled down the sample in zero field and apply the field $H$ just before the measurements, and we kept $H$ at a constant value during the data collecting. We obtained XMCD data by averaging 8 points with turning on the $H$ field, of which we altered the direction.

.

## RESULTS AND DISCUSSIONS

**Magnetization on different substrates**

Figure 4 summarizes the structural and magnetic properties of 30-nm SRO113 films on various substrates as a function of in-plane strain. In Figure 4(a), the out-of-plane *c*-axis lattice parameter and the unit-cell volume in the pseudo-cubic notations are plotted. As expected[7], compressive (expansive) in-plane strain leads to expansion (contraction) in the *c* axis, as well as the contraction (expansion) of the unit-cell volume. The change in the unit-cell volume corresponds to the biaxial Poisson ratio of $\nu_c$ = 0.62. Here $\nu_c$ = 2 corresponds to deformation satisfying the volume conservation[28].

Note that the values for the SRO214 substrate follow well with the variations for other pseudo-cubic perovskite substrates. However, the magnetization in Figure 4(b) clearly indicates the anomalous enhancement only for the SRO214 substrate. This fact reveals that the substrate-induced strain is not the direct origin of the enhanced *M*.

In Figure 4(c), $T_{\mathrm{Curie}}$ based on the *M*-*T* curves in Figure 3(a) is plotted. With NGO and STO substrates, $T_{\mathrm{Curie}}$ is substantially lower than that of the bulk SRO113. Such depression of $T_{\mathrm{Curie}}$ has been reported for strained SRO113[6]. In case of the DSO substrate, the film suffers the least strain among the substrates chosen in this study, but nevertheless it is well strained expansively with the substrate. The observed $T_{\mathrm{Curie}}$ of 160 K matches well with that of the bulk SRO113. Compared with the depressed $T_{\mathrm{Curie}}$ for STO with comparable but compressive biaxial strain, this contrast behavior implies certain anisotropy of magnetic properties with respect to the sign of the strain accompanied by the asymmetric volume change. For a GSO substrate, a major part of the film exhibits a similar depression in $T_{\mathrm{Curie}}$, but there is an additional part showing the $T_{\mathrm{Curie}}$ similar to that of the bulk SRO113, as shown in Figure 3(a) as a double-step transition. Such a separation of the transition is attributable to the separation of the film properties between a strained part near the interface and an unstrained part near the top surface, manifesting a particularly large expansive strain of the SRO113 film near the interface to GSO.

The importance of local magnetic moment in the SRO113 ferromagnetism was stressed by angle resolved photoemission spectroscopy measurement on an *in situ* grown SRO113 film[29]. Ferromagnetic resonance experiment on SRO113 was consistent with the existence of strong spin-orbit coupling[30]. In addition, strong Ru 4*d* – O 2*p* orbitals hybridization is found in a variety of experiments such as: the magnetic Compton profile[9], the optical conductivity spectra[10], and resonant photoemission spectroscopy[31].

Among various systems investigated in this study, SRO113/SRO214 exhibits exceptional magnetic behavior as shown in Figures 3 and 4(b). It has almost twice larger *M* than that of the other SRO113 films on insulating perovskite substrates and unambiguously exceeds its upper limit value of 2 $\mu_B/Ru^{4+}$. Interestingly, $T_{Curie}$ also shows the highest value even though SRO113 films are under severe compressive strain. Note, when we estimated *M* of each film, we carefully subtracted *M* values of substrates as described in **MATERIALS AND METHODS**. It is also observed that heating of substrate during SRO113 film deposition is not changing the magnetic behavior of SRO214 substrates (see Supplementary Fig. S2). This control experiment makes it clear that *M* of SRO113/SRO214 films is peculiarly large among all reported SRO113 film systems. To explain this difference, we further conducted a number of experiments.

**Origin of the magnetization enhancement in SRO113/SRO214**

From the discussion in the earlier section, strain-induced $RuO_6$ distortion is clearly ruled out as the origin of the enhanced magnetization in SRO113/SRO214. Considering SRO214 single crystal growth process, invasion of a small amount of SRO113 phase can be feasible in SRO214 substrates. As described in the **MATERIALS AND METHODS**, we always measured *M* of SRO214 substrates before SRO113 ablation to remove the magnetic signals from the substrate itself (Figure S1). Therefore, despite possible formation of SRO113 phase in SRO214, the enhanced magnetization of SRO113/SRO214 cannot be explained by the substrate contribution.

Because PLD growth is a dynamical process, every individual environment inside and outside the growth vacuum chamber can affect the film quality. We already suggested that the high crystallinity of all SRO113 films used in the present study. The high crystallinity is also supported by excellent electrical properties presented in Figure S1(d) with high residual resistance ratio (RRR) values. To reduce potential formation of oxygen vacancies in the film during the PLD process, especially during increasing substrate temperature up to the film growth temperature (700°C), the chamber was always filled with 0.1 Torr $O_2$ at all temperatures.

This process was tested again and the results were presented in Figures S3). Thus, effects of non-stoichiometry in both SRO113 films and SRO214 substrates are excluded as an origin of the enhanced *M*.

Let us turn to in-depth discussion of the magnetization enhancement of SRO113/SRO214. There were few reports on SRO113 films whose magnetization exceeds 2 $\mu_B/Ru^{4+}$ attributed to its high-spin configuration. Recent work on SRO113 films grown along the (111) direction on a STO(111) substrate showed enhanced *M* values[16]. The authors suggested possible existence of high-spin configuration in SRO113 by employing XAS and XMCD measurements. Nevertheless, for the SRO113 film grown along (001) direction on STO(001) they obtained magnetization results similar to ours (Figure 4(b)). Note that recent work on high-quality single crystalline SRO113 reports $T_{\mathrm{Curie}}$ = 163 K with *M* of ~1.0 and ~1.2 $\mu_B/Ru^{4+}$ along the pseudo-cubic *c*- and *a*-axes in the same measurement conditions as those for the SRO113 films on various substrates[32].

We investigated the electronic structure of the SRO113 films by Ru *M*-edge (Ru $3p \rightarrow 4d$) XAS measurements. Figure 5(a) shows Ru $M_{2,3}$-edge XAS spectra of SRO113 thin films on SRO214 substrate and a SRO214 bare substrate, as well as of bulk polycrystalline SRO113. Photon helicities were parallel ($\mu^+$) and antiparallel ($\mu^-$) to the majority spin direction of $Ru^{4+}$ set by the external field on cooling. The absorption peaks at 464.5 eV and 487 eV are owing to transitions from Ru $3_{p3/2}$ and $3_{p1/2}$ core levels to the Ru 4*d* bands. Additional shoulder peaks at 478.3 eV and 500.5 eV are due to transitions to Ru 5*s* bands. These peak assignments are consistent with the previous bulk SRO113 XAS spectra[33]. Note that these peaks are also similar to those of bare SRO214 substrate (See the bottom curve in Figure 5(a)). Although most ruthenium oxides have $Ru^{4+}$ valence states with the low-spin configuration, there are a number of examples in which changes of spin configuration or valence states result in large changes in XAS spectra[34, 35]. Our observation here confirms that the films have $Ru^{4+}$ ions in the low-spin state.

XMCD measures difference in response from photon helicities: i.e. parallel ($\mu^{+}$) and antiparallel ($\mu^{-}$) to the majority spin direction of $Ru^{4+}$. Without ferromagnetism, there should be no XMCD signal. As shown in Figure 5(b), the bare SRO214 indeed does not show any XMCD signal. In contrast, the SRO113 films show clear XMCD signals, indicating that there are ferromagnetic alignments in Ru-4*d* electrons at the measurement temperature of 100 K. The peak separation in Figure 5(b) agree with that of the bulk SRO113, which is assigned as the energy difference between Ru $3_{p3/2}$ and $3_{p1/2}$[36]. Therefore, the electronic states of the films on SRO214 are very close to those of the bulk SRO113; it is unlikely that that the high-spin configuration is realized in the thin films on SRO214 substrates.

Contrary to the case of Ru ions, the oxygen ions in SRO113 and SRO214 can have quite different environments. Namely, SRO214 have distinct apical and in-plane oxygens due to its layered crystal structure. Figure 5(c) shows O *K*-edge (O 1*s* $\rightarrow$ 2*p*) XAS spectra of the SRO113 films on SRO substrates and of a SRO214 bare substrate. Note that the XAS spectra of SRO214 have two peaks in the energy region between 529 and 531 eV, which can be assigned as the transition from apical oxygen bands to Ru 4*d*-*yz*/*zx* and from in-plane oxygen to 4*d*-*xy* states[33]. These spectral features are different from the XAS spectra of the SRO113 thin films. This difference might provide a way to look into the possibility of induced spin-polarization in the SRO214 substrate beneath the SRO113 layer.

Since the X-ray penetrates into the substrate through the film, the substrate contribution can be seen in the photon energy region (i.e. below 529.5 eV) where the absorption of the film is small. It is noticeable that the 1.2 nm film has a weak peak around 529.3 eV in the XAS spectrum, indicating the contribution from the substrate. We also obtained the XMCD spectra of 1.2 nm and 30 nm SRO113 films as shown in Figure 5(d). They look nearly the same as those of the bulk SRO113[36] and do not exhibit any evident peak near 529.3 eV. To clarify this point, we normalized the XMCD spectra of the 1.2 nm and 30 nm SRO113 films with their peak intensities and plotted them in the inset of Figure 5(d). These normalized XMCD spectra are also very close to that of a polycrystalline SRO113. These results suggest that, the spin polarization in the *yz*/*zx* orbitals of the SRO214 substrate, if any, are quite small.

Our systematic analysis presented leads us to look into the interfacial state between SRO113 and SRO214, which has not been discussed yet. SRO214 is clearly distinct from other substrates used: it is the only metallic substrate with similar $Ru^{4+}$ electronic configurations to SRO113. With the SRO214 substrate, ferromagnetic ordering with $T_{Curie}$~ 160 K appears even with a thickness as small as 1.2 nm, whereas with the STO substrate, for example, it is known that ferromagnetic order is strongly suppressed with the film thickness below 3 uc[37]. SRO214 is a paramagnetic metal and a spin-triplet superconductor below $T_c$=1.5 K. When SRO214 forms a good metallic interface with a ferromagnet, a magnetic moment may be induced in SRO214 near its interface. Insulating STO single crystal turns to metallic with an introduction of a dopant such as La and Nb, but is known to form a Schottky-like interfacial barrier with a SRO113 film. When a conducting substrate of 0.5 wt.% Nb-doped STO (Nb:STO) was used, *M* of SRO113/Nb:STO is smaller than that using an undoped insulating STO substrate (Figure S4). This result indicates that a conducting substrate with a non-Ohmic contact with ferromagnetic SRO113 has little influence on the additional magnetic moment induction in SRO113. In contrast with SRO113/Nb:STO, we demonstrated an excellent metallic contact between SRO113 and SRO214 in our earlier report[23], which may easily induce ferromagnetic moment on the paramagnetic SRO214 metal side. The common electronic character with $Ru^{4+}$ moment may also promote the magnetic moment induction across the interface. However, direct measurements of the induced magnetic moment at the buried SRO113/SRO214 interface are not easy. We note that superlattices of SRO113 with the ferromagnetic metal $La_{0.7}Sr_{0.3}MnO_3$ (LSMO) shows enhanced magnetization compared to their single material layers[2]. The enhancement is attributed to an antiferromagnetic coupling emerging at the LSMO/SRO113 interface. Similarly, it is a feasible possibility that magnetization enhancement at ferromagnet/paramagnet hetero-interface of SRO113/SRO214 originates from yet unknown magnetic coupling.

Figure 6 characterizes the amount of enhancement as a function of the SRO113 film thickness. The solid red line is a linear fit to the *M* after careful subtraction of the background substrate-only contributions (shown by blue circles) by a procedure described in **MATERIALS AND METHODS**. The blue line is the anticipated SRO113 film contribution with a known bulk *M* of SRO113[38]: 1.8 $\mu_B/Ru^{4+}$. The difference between the red and blue lines corresponds to enhancement over the anticipated contribution from the SRO113 film. It does not depend much on the film thickness. Assuming that the magnetic moment is induced in SRO214 near the SRO113/SRO214 interface, contribution of the induced magnetic moment from SRO214 and SRO113 film only is distinguished with dark and light grey colors, respectively.

## CONCLUSIONS

In this study, we found an exceptionally enhanced magnetization in SRO113/SRO214. This enhancement is neither due to strain-induced structural distortion nor due to substrate metallicity alone. Electronic structure change, from the low-spin to high-spin configuration, is not observed either. The film-thickness dependence of the observed magnetization systematically indicates an additional contribution that depends only weakly on the thickness. Therefore, we conclude that an induced magnetic moment in SRO214 near the SRO113/SRO214 interface be responsible for the additional magnetism in the SRO113/SRO214 system. Further confirmation of the spatial distribution of the induce magnetism at the SRO113/SRO214 interface as well as the clarification of its mechanism is needed in future.

## ACKNOWLEDGEMENTS

This work was supported by the Research Center Program of IBS (Institute for Basic Science) in Korea (IBS-R009-D1). It was also supported by the JSPS KAKENHI Grant No. JP22103002 (Topological Quantum Phenomena) and No. JP15H05852 (Topological Materials Science). MSA is supported as an International Research Fellow of the Japan Society for the Promotion of Science (JSPS).

**Competing Interests** The authors declare that they have no competing financial interests.

**Corresponding author:** M. S. Anwar, Department of Physics, Graduate School of Science, Kyoto University, Kyoto 606-8502, Japan. E-mail: anwar@scphys.kyoto-u.ac.jp

*S. R. Lee. Present address: Max Planck POSTECH/Korea Research Initiative, Pohang, Gyeongbuk 790-784, Republic of Korea. E-mail: leeseungran@gmail.com

## References

1. Koster, G., Klein, L., Siemons, W., Rijnders, G., Dodge, J. S., Eom, C.-B., Blank, D. H. A. and Beasley, M. R. Structure, physical properties, and applications of $SrRuO_3$ thin films. **84**, 253-298 *Rev. Mod. Phys.* (2012).

2. Ziese, M., Vrejoiu, I., Pippel, E., Esquinazi, P., Hesse, D., Etz, C., Henk, J., Ernst, A., Maznichenko, I. V., Hergert, W. and Mertig, I. Tailoring Magnetic Interlayer Coupling in $La_{0.7}Sr_{0.3}MnO_3/SrRuO_3$ Superlattices. **104**, 167203 *Phys. Rev. Lett.* (2010).

3. Ahn, C. H., Tybell, T., Antognazza, L., Char, K., Hammond, R. H., Beasley, M. R., Fischer, Ø. and Triscone, J. M. Local, Nonvolatile Electronic Writing of Epitaxial $Pb(Zr_{0.52}Ti_{0.48})O_3/SrRuO_3$ Heterostructures. **276**, 1100-1103 *Science* (1997).

4. Catalan, G., Lubk, A., Vlooswijk, A. H. G., Snoeck, E., Magen, C., Janssens, A., Rispens, G., Rijnders, G., Blank, D. H. A. and Noheda, B. Flexoelectric rotation of polarization in ferroelectric thin films. **10**, 963-967 *Nat. Mater.* (2011).

5. Antognazza, L., Char, K., Geballe, T. H., King, L. L. H. and Sleight, A. W. Josephson coupling of $YBa_2Cu_3O_{7-x}$ through a ferromagnetic barrier $SrRuO_3$. **63**, 1005-1007 *Appl. Phys. Lett.* (1993).

6. Gan, Q., Rao, R. A., Eom, C. B., Garrett, J. L. and Lee, M. Direct measurement of strain effects on magnetic and electrical properties of epitaxial $SrRuO_3$ thin films. **72**, 978-980 *Appl. Phys. Lett.* (1998).

7. Zayak, A. T., Huang, X., Neaton, J. B. and Rabe, K. M. Structural, electronic, and magnetic properties of $SrRuO_3$ under epitaxial strain. **74**, 094104 *Phys. Rev. B* (2006).

8. Gunnarsson, R. Anisotropic spin-orbit interaction revealed by in-plane magnetoresistance in single-oriented $SrRuO_3$ thin films. **85**, 235409 *Phys. Rev. B* (2012).

9. Hiraoka, N., Itou, M., Deb, A., Sakurai, Y., Kakutani, Y., Koizumi, A., Sakai, N., Uzuhara, S., Miyaki, S., Koizumi, H., Makoshi, K., Kikugawa, N. and Maeno, Y. Ru-O orbital hybridization and orbital occupation in $SrRuO_3$: A magnetic Compton-profile study. **70**, 054420 *Phys. Rev. B* (2004).

10. Lee, J. S., Lee, Y. S., Noh, T. W., Nakatsuji, S., Fukazawa, H., Perry, R. S., Maeno, Y., Yoshida, Y., Ikeda, S. I., Yu, J. and Eom, C. B. Bond-length dependence of charge-transfer excitations and stretch phonon modes in perovskite ruthenates: Evidence of strong *p-d* hybridization effects. **70**, 085103 *Phys. Rev. B* (2004).

11. Herranz, G., Martínez, B., Fontcuberta, J., Sánchez, F., Ferrater, C., García-Cuenca, M. V. and Varela, M. Enhanced electron-electron correlations in nanometric $SrRuO_3$ epitaxial films. **67**, 174423 *Phys. Rev. B* (2003).

12. Guedes, E. B., Abbate, M., Ishigami, K., Fujimori, A., Yoshimatsu, K., Kumigashira, H., Oshima, M., Vicentin, F. C., Fonseca, P. T. and Mossanek, R. J. O. Core level and valence band spectroscopy of $SrRuO_3$: Electron correlation and covalence effects. **86**, 235127 *Phys. Rev B.* (2012).

13. Zayak, A. T., Huang, X., Neaton, J. B. and Rabe, K. M. Manipulating magnetic properties of $SrRuO_3$ and $CaRuO_3$ with epitaxial and uniaxial strains. **77**, 214410 *Phys. Rev. B* (2008).

14. Cheng, J.-G., Zhou, J.-S. and Goodenough, J. B. Lattice effects on ferromagnetism in perovskite ruthenates. **110**, 13312-13315 *Proc. Natl. Acad. Sci.* (2013).

15. Eom, C. B., Cava, R. J., Fleming, R. M., Phillips, J. M., vanDover, R. B., Marshall, J. H., Hsu, J. W. P., Krajewski, J. J. and Peck, W. F. Single-Crystal Epitaxial Thin Films of the Isotropic Metallic Oxides $Sr_{1-x}Ca_xRuO_3$ ($0 \le x \le 1$). **258**, 1766-1769 *Science* (1992).

16. Grutter, A. J., Wong, F. J., Arenholz, E., Vailionis, A. and Suzuki, Y. Evidence of high-spin Ru and universal magnetic anisotropy in $SrRuO_3$ thin films. **85**, 134429 *Phys. Rev. B* (2012).

17. Kim, K. W., Lee, J. S., Noh, T. W., Lee, S. R. and Char, K. Metal-insulator transition in a disordered and correlated $SrTi_{1-x}Ru_xO_3$ system: Changes in transport properties, optical spectra, and electronic structure. **71**, 125104 *Phys. Rev. B* (2005).

18. Grutter, A. J., Wong, F. J., Jenkins, C. A., Arenholz, E., Vailionis, A. and Suzuki, Y. Stabilization of spin-zero $Ru^{4+}$ through epitaxial strain in $SrRuO_3$ thin films. **88**, 214410 *Phys. Rev. B* (2013).

19. Maeno, Y., Hashimoto, H., Yoshida, K., Nishizaki, S., Fujita, T., Bednorz, J. G. and Lichtenberg, F. Superconductivity in a layered perovskite without copper. **372**, 532-534 *Nature* (1994).

20. Mackenzie, A. P. and Maeno, Y. The superconductivity of $Sr_2RuO_4$ and the physics of spin-triplet pairing. **75**, 657-712 *Rev. Mod. Phys.* (2003).

21. Maeno, Y., Kittaka, S., Nomura, T., Yonezawa, S. and Ishida, K. Evaluation of Spin-Triplet Superconductivity in $Sr_2RuO_4$. **81**, 011009 *J. Phys. Soci. Jpn.* (2011).

22. Anwar, M. S. *et al.* Direct penetration of spin-triplet superconductivity into a ferromagnet in Au/$SrRuO_3$/$Sr_2RuO_4$ junctions. *Nat. Commun.* (accepted). arXiv:1603.00971v1

23. Anwar, M. S., Shin, Y. J., Lee, S. R., Kang, S. J., Sugimoto, Y., Yonezawa, S., Noh, H.-J. and Maeno, Y. Ferromagnetic $SrRuO_3$ thin-film deposition on a spin-triplet superconductor $Sr_2RuO_4$ with a highly conducting interface. **8**, 019202 *Appl. Phys. Express* (2015).

24. Gillot, C., Michenaud, J. P., Maglione, M. and Jannot, B. DC electrical resistivity of Nb-Doped $BaTiO_3$ and EPR measurements. **84**, 1033-1038 *Solid State Commun.* (1992).

25. Matzdorf, R., Fang, Z., Ismail, Zhang, J., Kimura, T., Tokura, Y., Terakura, K. and Plummer, E. W. Ferromagnetism Stabilized by Lattice Distortion at the Surface of the p-Wave Superconductor $Sr_2RuO_4$. **289**, 746-748 *Science* (2000).

26. Firmo, I. A., Lederer, S., Lupien, C., Mackenzie, A. P., Davis, J. C. and Kivelson, S. A. Evidence from tunneling spectroscopy for a quasi-one-dimensional origin of superconductivity in $Sr_2RuO_4$. **88**, 134521 *Phys. Rev. B* (2013).

27. Sugimoto, Y., Anwar, M. S., Lee, S. R., Shin, Y. J., Yonezawa, S., Noh, T. W. and Maeno, Y. Ferromagnetic Properties of $SrRuO_3$ Thin Films Deposited on the Spin-Triplet Superconductor $Sr_2RuO_4$. **75**, 413-418 *Physics Procedia* (2015).

28. Hammerschmidt, T., Kratzer, P. and Scheffler, M. Elastic response of cubic crystals to biaxial strain: Analytic results and comparison to density functional theory for InAs. **75**, 235328 *Phys. Rev. B* (2007).

29. Shai, D. E., Adamo, C., Shen, D. W., Brooks, C. M., Harter, J. W., Monkman, E. J., Burganov, B., Schlom, D. G. and Shen, K. M. Quasiparticle Mass Enhancement and Temperature Dependence of the Electronic Structure of Ferromagnetic $SrRuO_3$ Thin Films. **110**, 087004 *Phys. Rev. Lett.* (2013).

30. Langner, M. C., Kantner, C. L. S., Chu, Y. H., Martin, L. M., Yu, P., Seidel, J., Ramesh, R. and Orenstein, J. Observation of Ferromagnetic Resonance in $SrRuO_3$ by the Time-Resolved Magneto-Optical Kerr Effect. **102**, 177601 *Phys. Rev. Lett.* (2009).

31. Grebinskij, S., Senulis, M., Tvardauskas, H., Bondarenka, V., Lisauskas, V., Vengalis, B., Orlowski, B. A., Johnson, R. L. and Mickevičius, S. Electronic structure of epitaxial $SrRuO_3$ films studied by resonant photoemission. **80**, 1140-1144 *Rad. Phys. Chem.* (2011).

32. Kikugawa, N., Baumbach, R., Brooks, J. S., Terashima, T., Uji, S. and Maeno, Y. Single-Crystal Growth of a Perovskite Ruthenate $SrRuO_3$ by the Floating-Zone Method. **15**, 5573-5577 *Crystal Growth & Design* (2015).

33. Noh, H.-J., Oh, S. J., Park, B. G., Park, J. H., Kim, J. Y., Kim, H. D., Mizokawa, T., Tjeng, L. H., Lin, H. J., Chen, C. T., Schuppler, S., Nakatsuji, S., Fukazawa, H. and Maeno, Y. Electronic structure and evolution of the orbital state in metallic $Ca_{2-x}Sr_xRuO_4$. **72**, 052411 *Phys. Rev. B* (2005).

34. Haverkort, M. W., Hu, Z., Cezar, J. C., Burnus, T., Hartmann, H., Reuther, M., Zobel, C., Lorenz, T., Tanaka, A., Brookes, N. B., Hsieh, H. H., Lin, H. J., Chen, C. T. and Tjeng, L. H. Spin State Transition in $LaCoO_3$ Studied Using Soft X-ray Absorption Spectroscopy and Magnetic Circular Dichroism. **97**, 176405 *Phys. Rev. Lett.* (2006).

35. Agrestini, S., Hu, Z., Kuo, C. Y., Haverkort, M. W., Ko, K. T., Hollmann, N., Liu, Q., Pellegrin, E., Valvidares, M., Herrero-Martin, J., Gargiani, P., Gegenwart, P., Schneider, M., Esser, S., Tanaka, A., Komarek, A. C. and Tjeng, L. H. Electronic and spin states of $SrRuO_3$ thin films: An x-ray magnetic circular dichroism study. **91**, 075127 *Phys. Rev. B* (2015).

36. Okamoto, J., Okane, T., Saitoh, Y., Terai, K., Fujimori, S. I., Muramatsu, Y., Yoshii, K., Mamiya, K., Koide, T., Fujimori, A., Fang, Z., Takeda, Y. and Takano, M. Soft x-ray magnetic circular dichroism study of $Ca_{1-x}Sr_xRuO_3$ across the ferromagnetic quantum phase transition. **76**, 184441 *Phys. Rev. B* (2007).

37. Chang, Y. J., Kim, C. H., Phark, S. H., Kim, Y. S., Yu, J. and Noh, T. W. Fundamental Thickness Limit of Itinerant Ferromagnetic $SrRuO_3$ Thin Films. **103**, 057201 *Phys. Rev. Lett.* (2009).

38. Bern, F., Ziese, M., Setzer, A., Pippel, E., Hesse, D. and Vrejoiu, I. Structural, magnetic and electrical properties of $SrRuO_3$ films and $SrRuO_3/SrTiO_3$ superlattices. **25**, 496003 *J. Phys.: Cond. Matter* (2013).

## Figure Captions

**Figure 1** (**a**) X-ray diffraction spectra of 30-nm thick $SrRuO_3$ films grown on various substrates. Sharper peaks correspond to the substrate peaks: (004) and (006) for $Sr_2RuO_4$ and (001) and (002) for the others. The dashed lines indicate the positions of the (001) and (002) peaks of bulk $SrRuO_3$. (**b**) Rocking curves around the $SrRuO_3$ (002) peak.

**Figure 2** X-ray reciprocal space mapping of 30-nm $SrRuO_3$ films grown on (**a**) $NdGaO_3$, (**b**) $SrTiO_3$, (**c**) $DyScO_3$, and (**d**) $GdScO_3$ substrates around asymmetric (103) Bragg reflection of all substrates in the cubic and pseudocubic structures. The axes correspond to the inverse of the in-plane and out-of-plane lattice constants of each material. The vertical dashed lines are guides to demonstrate that the films are fully strained on the substrates. Thickness fringes of SRO113 films are clearly seen in (b) and (c).

**Figure 3** Temperature dependence of remnant magnetization of 30-nm thick $SrRuO_3$ thin films after fields along (a) the *c*-axis and (b) the *a*-axis are turned off. The substrate contribution has been subtracted. The films were cooled down to 4 K under the applied field of 1 T and the magnetization data were collected during warming under zero field. The insets show the details around the Curie temperature.

**Figure 4** Structural and magnetic properties of $SrRuO_3$ thin films under in-plane strain. (**a**) The out-of-plane lattice parameters (black filled square, left axis) and the unit-cell volumes (red filled square, right axis), in the orthorhombic structure evaluated from the X-ray diffraction data. The horizontal dashed lines indicate the unstrained bulk properties of $SrRuO_3$. (**b**) Magnetization per Ru atom in the film and (**c**) the Curie temperatures. The horizontal solid lines in (**b** and **c**) are representing the values corresponding for single crystal and thin films reported in ref. 31 (dashed line) and ref. 37 (solid line), respectively.

**Figure 5** XAS and XMCD spectra of 1.2-nm (blue) and 30-nm (red) $SrRuO_3$ films on $Sr_2RuO_4$ substrate, as well as of $Sr_2RuO_4$ substrate alone (black) at Ru *M*-edges (**a, b**) and O *K*-edge (**c, d**) at 100 K. (**a**) XAS intensities with the positive helicity $\mu^+$ (solid lines) and negative helicity $\mu^-$ (dotted lines) and (**b**) XMCD spectral intensity $\Delta\mu = \mu^+ - \mu^-$. $SrRuO_3$ bulk XAS spectrum (green) is taken from ref. 34. The photon energy value is calibrated by the peak differences between XAS and XMCD data. (**c**) XAS and (**d**) XMCD spectra at O *K*-edge. The solid triangles and labels represent origin of the transitions by the positions and characters of unoccupied *d* bands. The inset in (**d**) shows normalized (with the peak intensity) data of thin films and polycrystalline $SrRuO_3$, taken from ref. 34.

**Figure 6** Schematic of magnetization contributions in $SrRuO_3/Sr_2RuO_4$ with various values of film thickness. The measured magnetization of $Sr_2RuO_4$ substrates only and that of $SrRuO_3$ films on the $Sr_2RuO_4$ substrates are indicated by circles and squares, respectively. The solid red line shows a linear fit. The solid blue line indicates the magnetization based on the expected value of 1.9 $\mu_B$/Ru from the film (taken from ref. 37). The difference between the measured and expected values is indicated by light gray area that corresponds to the induced magnetization, most probably in the $Sr_2RuO_4$ substrate.

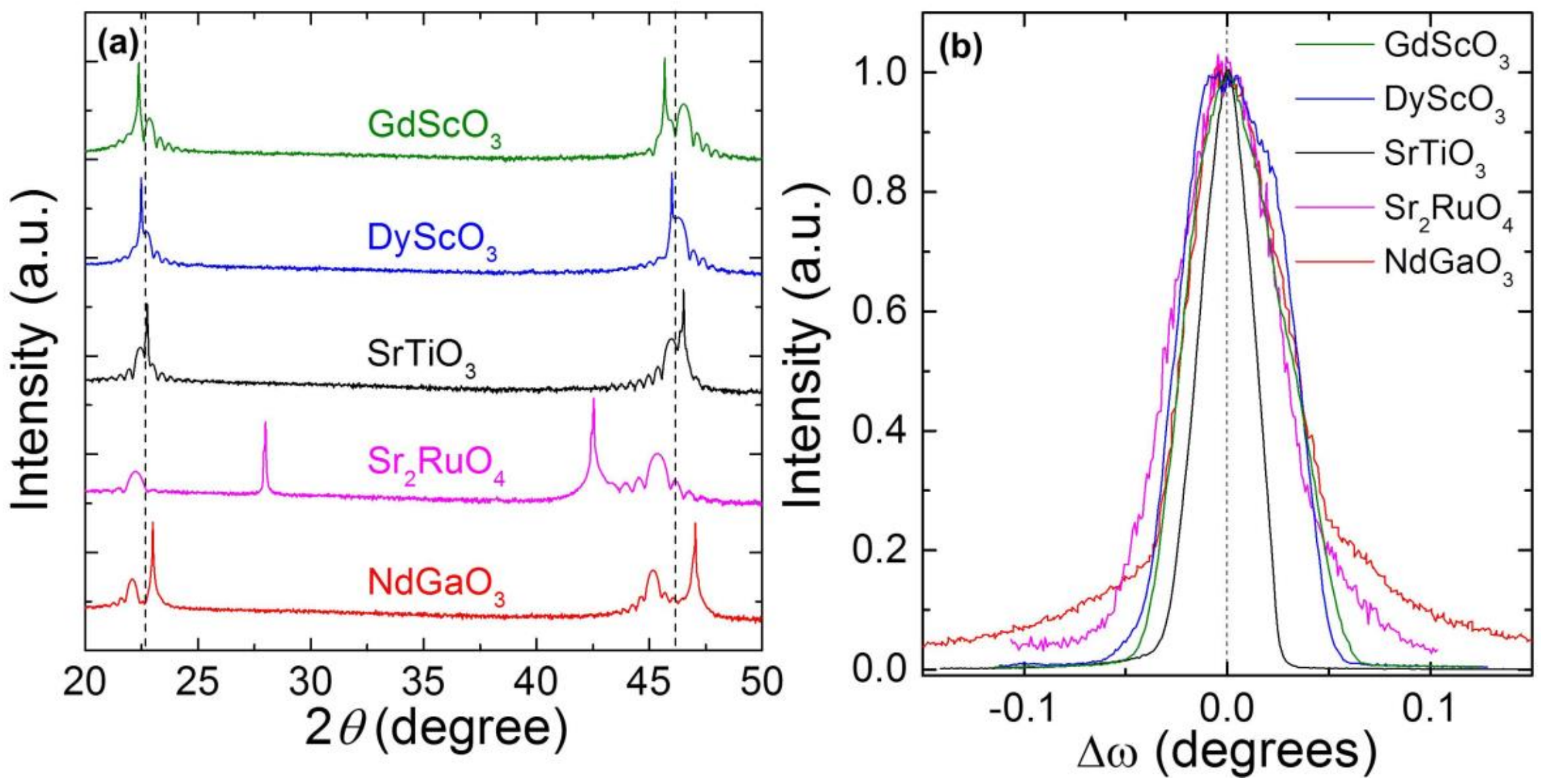

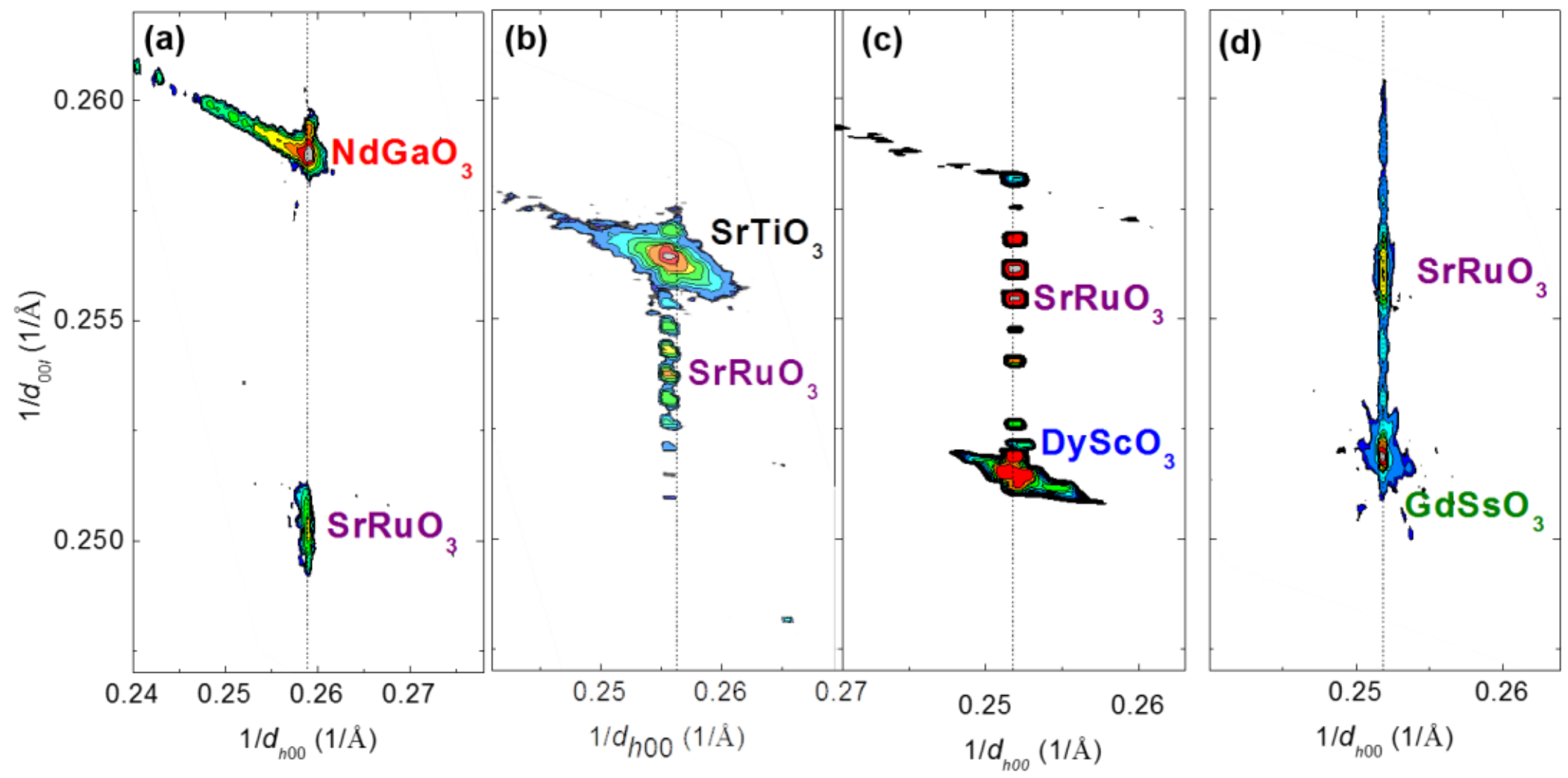

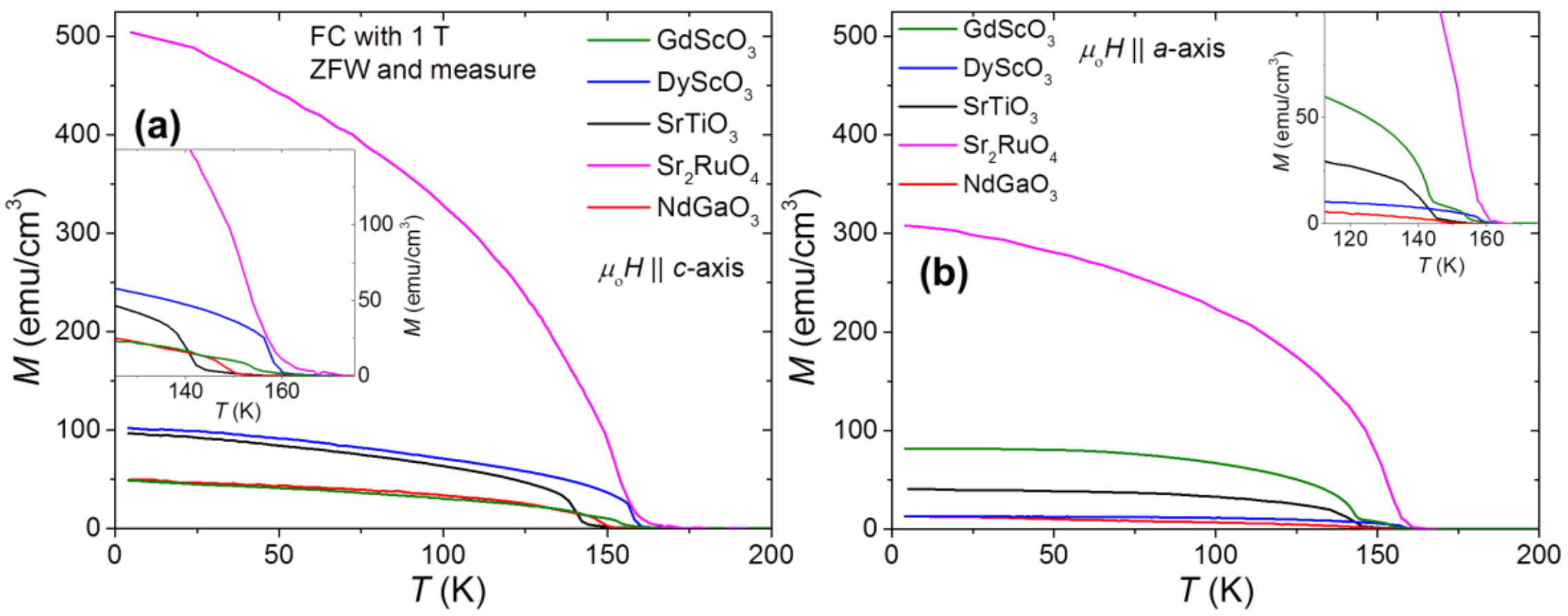

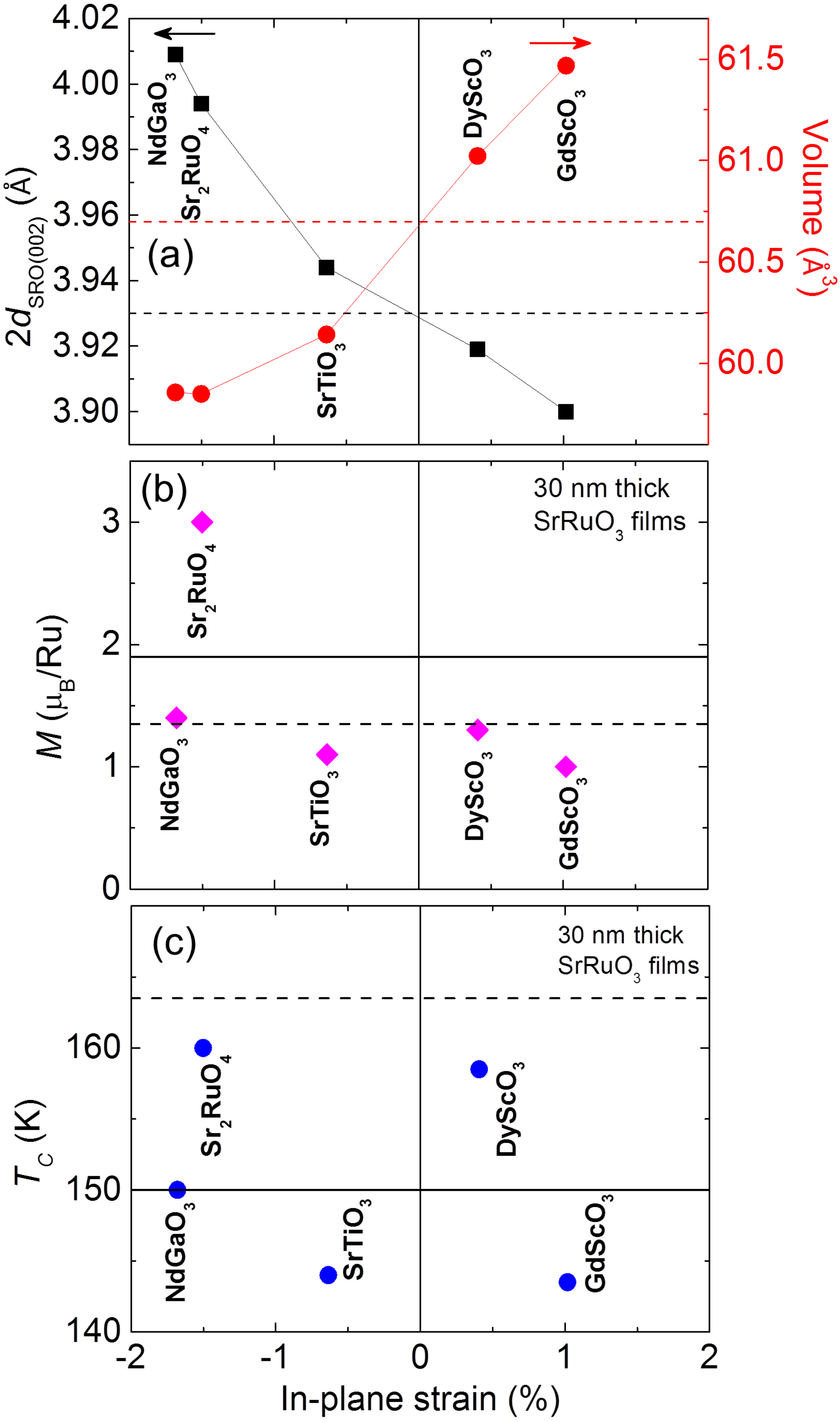

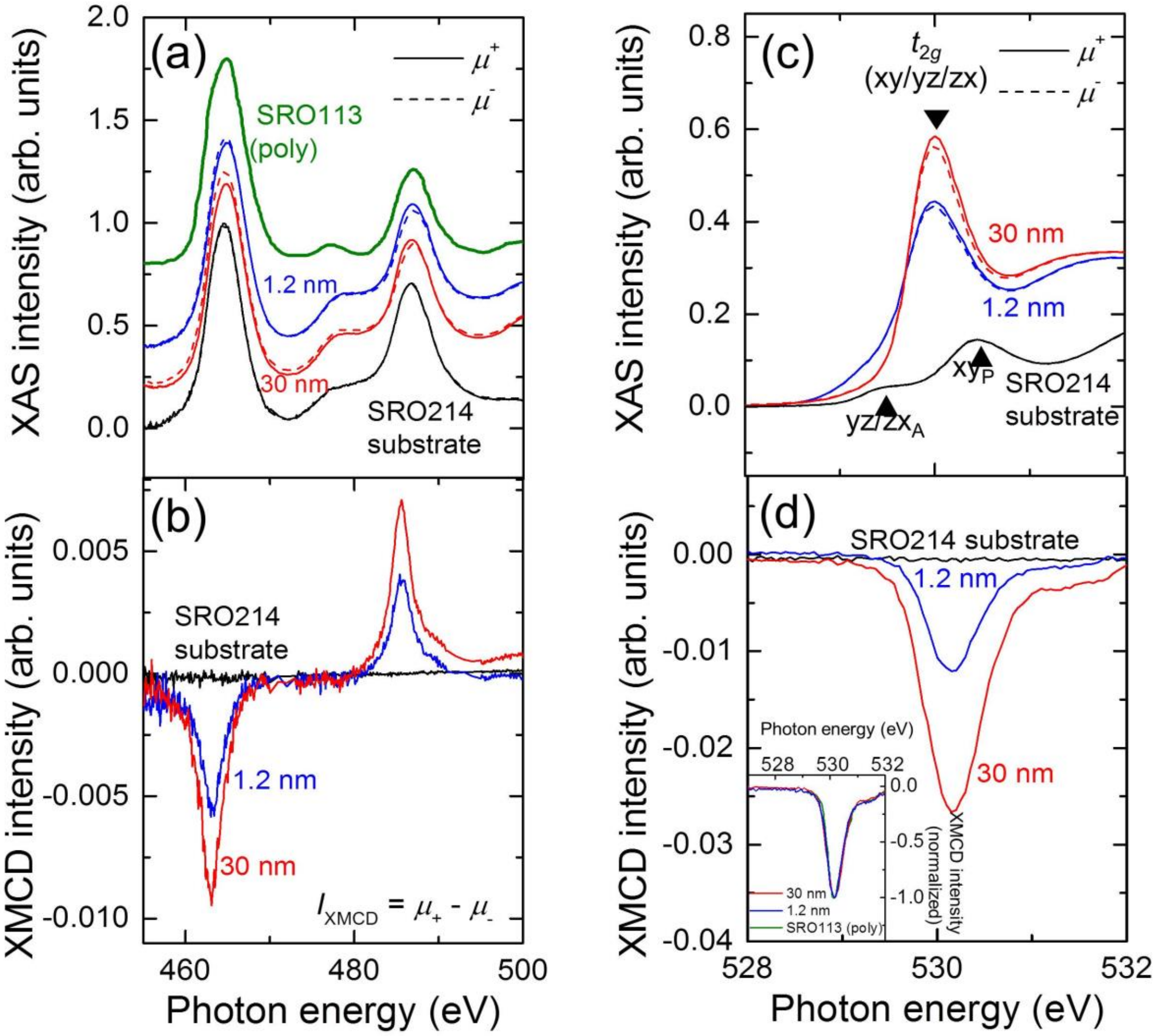

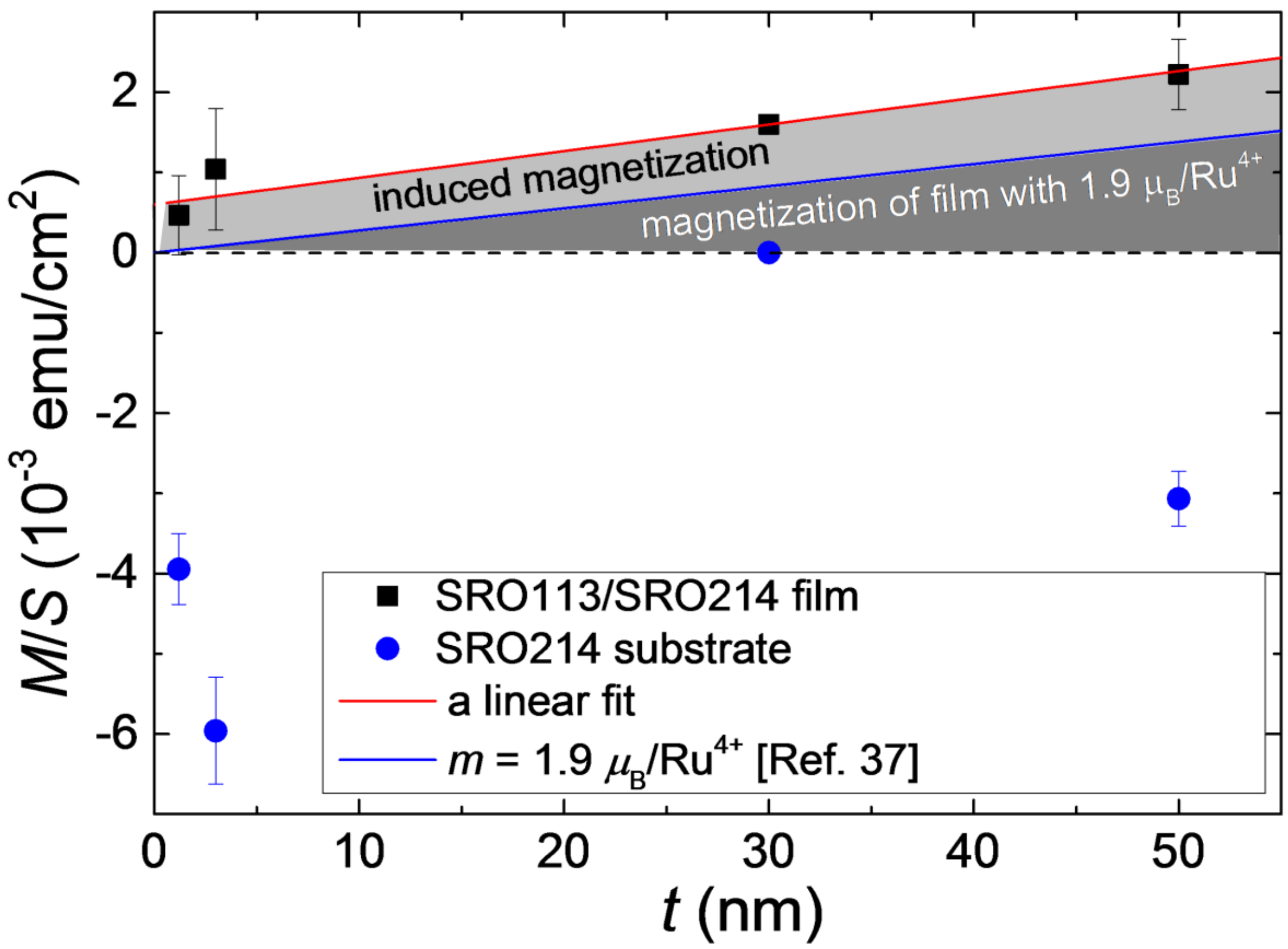

induced magnetization
magnetization of film with 1.9 μB/Ru4+
SRO113/SRO214 film
SRO214 substrate
a linear fit
m = 1.9 μB/Ru4+ [Ref. 37]
M/S (10-3 emu/cm2)
t (nm)

Supplementary information

# Induced ferromagnetic moment at the interface between epitaxial $SrRuO_3$ film and $Sr_2RuO_4$ single crystal

Seung Ran Lee[1,2], Muhammad Shahbaz Anwar[3], Yeong Jae Shin[1,2], Min-Cheol Lee[1,2],

Yusuke Sugimoto[3], Masanao Kunieda[3], Shingo Yonezawa[3], Yoshiteru Maeno[3]

& Tae Won Noh[1,2]

[1] Center for Correlated Electron Systems, Institute for Basic Science (IBS), Seoul 151-747, Republic of Korea

[2] Department of Physics and Astronomy, Seoul National University, Seoul 151-747, Republic of Korea

[3] Department of Physics, Graduate School of Science, Kyoto University, Kyoto 606-8502, Japan

Here, we are providing additional data on magnetization and resistivity. We present procedures of subtraction of SRO214 substrate contributions in magnetization values of SRO113/SRO214 thin films. We also present additional information on the SRO113 films deposited on various substrates including the temperature dependent resistance. Furthermore, we present the magnetization of a 30-nm thick SRO113 thin films deposited on Nb:STO substrate to study the effect of the conductivity of substrates on magnetic properties of the SRO113 film.

## 1. Subtraction of SRO214 substrate contribution of magnetization

We report unexpectedly high magnetization of SRO113/SRO214 thin films. To confirm such results, it is very important to carefully subtract all the contribution in magnetization other than SRO113 thin films. For this purpose, we subtracted the contributions of SRO214 substrates; furthermore, we investigated the effect of substrate-temperature on magnetization of the substrate during the deposition of SRO113 thin films. The details of both these procedures are given below.

## 1.1. Magnetization measurements before and after deposition of SRO113 thin films

To grow high quality SRO214 single crystals, about 15% of excess $RuO_2$ is added to compensate for the $RuO_2$ evaporation during melting. Due to higher contraction of $RuO_2$, other impurity phases such as ferromagnetic SRO113, paramagnetic $Sr_3Ru_2O_7$ (SRO327) and paramagnetic Ru metal may appear in the crystal. The density of Ru-inclusions can be controlled by using optimized concentration of $RuO_2$. Nevertheless, it is rather difficult to avoid SRO113 and SRO327 phases. Since we are using SRO214 as substrate, it is very important to quantify the amount of SRO113 impurities in the SRO214 substrate to obtain the accurate value of magnetization of SRO113 thin films. For this purpose, we measure the remnant magnetization of SRO214 substrate ($M_{214}$) using SQUID-magnetometer (Quantum design) before deposition of SRO113 thin films. It was measured on warming at zero field after field cooling at 1 T; the same procedure as described in MATERIALS AND METHODS of the main text. We detected the SRO113 impurity with the concentration varying from 5 ppm to 500 ppm depending on the crystal and also a location within the crystal. After deposition of SRO113 thin films of required thickness, magnetization ($M_{113/214}$) is measured again using the same magnetometer. To obtain the magnetization of SRO113 thin films ($M_{113}$), $M_{214}$ is subtracted from $M_{113/214}$ ($M_{113} = M_{113/214} - M_{214}$). An example of a set of date obtained in this procedure is presented in Figure S1. After such careful subtraction of substrate contributions, we still observe unexpectedly high magnetization of SRO113/SRO214 thin films, revealing that the enhanced magnetization originates from the thin films.

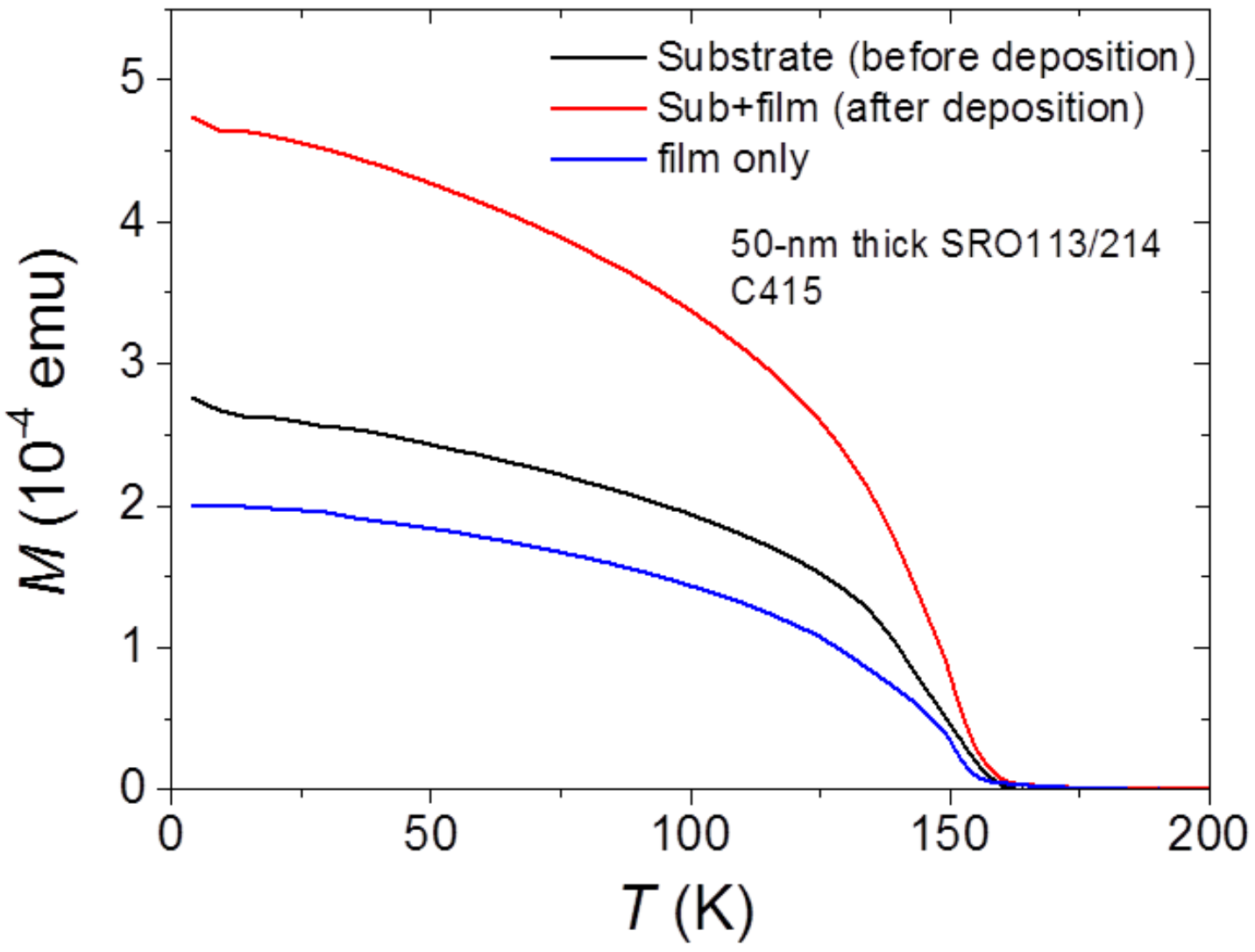

**Figure S1. Subtraction of substrate contribution of magnetization.** Remnant magnetization of SRO214 substrate after field cooling at 1 T was measured as a function of temperature before and after deposition of a SRO113 thin film.

### 1.2. Effect of substrate heating on magnetization

SRO113 thin films are grown at substrate-temperature of about 700°C. It has already been investigated that annealing process does not change the superconducting transition temperature of SRO214; it reflects the stability of SRO214 single crystals. Nevertheless, keeping the substrate at this high temperature for three hours may modify the magnetic impurities inside the SRO214 substrate. To investigate this possibility, we measured the magnetization of SRO214 substrate before and after the heating. We do not observe a significant difference in the magnetization that may affect our conclusion of the magnetization enhancement of the SRO113 films.

To heat up the substrate, we used the procedure shown in the inset of Figure S2(a), which simulates the actual deposition, except for the value of the background pressure. SRO214 substrates wereT heated up to 700°C in one hour in the vacuum with backing pressure of $10^{-6}$ Torr, stabilized the temperature at 700°C for one hour and finally cooled down to room temperature in one hour. The results shown in Figure S2 suggest that the estimated magnetization of the SRO113 film provide a lower limit and supports the conclusion that the enhanced magnetization arises because of the deposition of the SRO113 thin films.

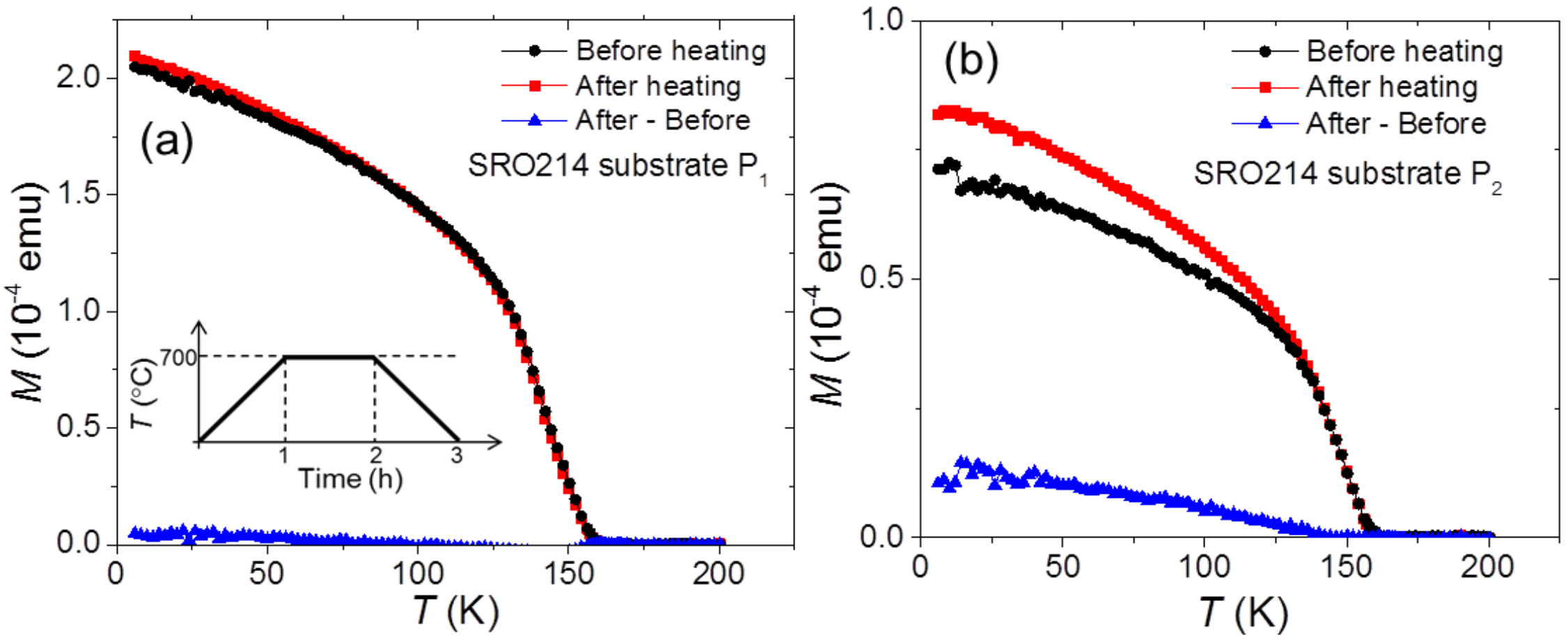

**Figure S2. Substrate heating effect on magnetization.** Magnetization of SRO214 substrates was measured before and after heating up to 700°C. For substrate C422-p1, there is a minute change in the magnetization (a). In contrast, there is about 15% increase in the magnetization after heating for another substrate C422-p2 (b). Note that both substrates are prepared from the same crystal. The inset of (a) illustrated the heating protocol. In both cases, the heating effect does not enhance the estimated contribution of the magnetization of the SRO113 film.

## 2. Film growth and resistance of SRO113 films on various substrates

Figure S3(a) shows the resistance as a function of temperature $R(T)$ of four 30-nm thin films of SRO113 deposited on STO, NGO, GSO and DSO substrates. $R(T)$ is measured using un-etched full $5 \times 5$ $mm^2$ films using a commercial system (Quantum Design, model PPMS) from 300 K to 4 K. The resistance decreases linearly down to $T_{Curie}$, below which a strong reduction in resistance occurs due to suppression of magnetic scattering. Figure S3(b) represents the temperature derivative d$R$/d$T$. The peak temperatures correspond well to $T_{Curie}$ determined from the magnetization $M(T)$. Residual resistance ratios (RRR) of these films are in the range of 5 – 7, assuring the high quality of these films. Table S1 summarizes the properties of these films deposited on a variety of substrates.

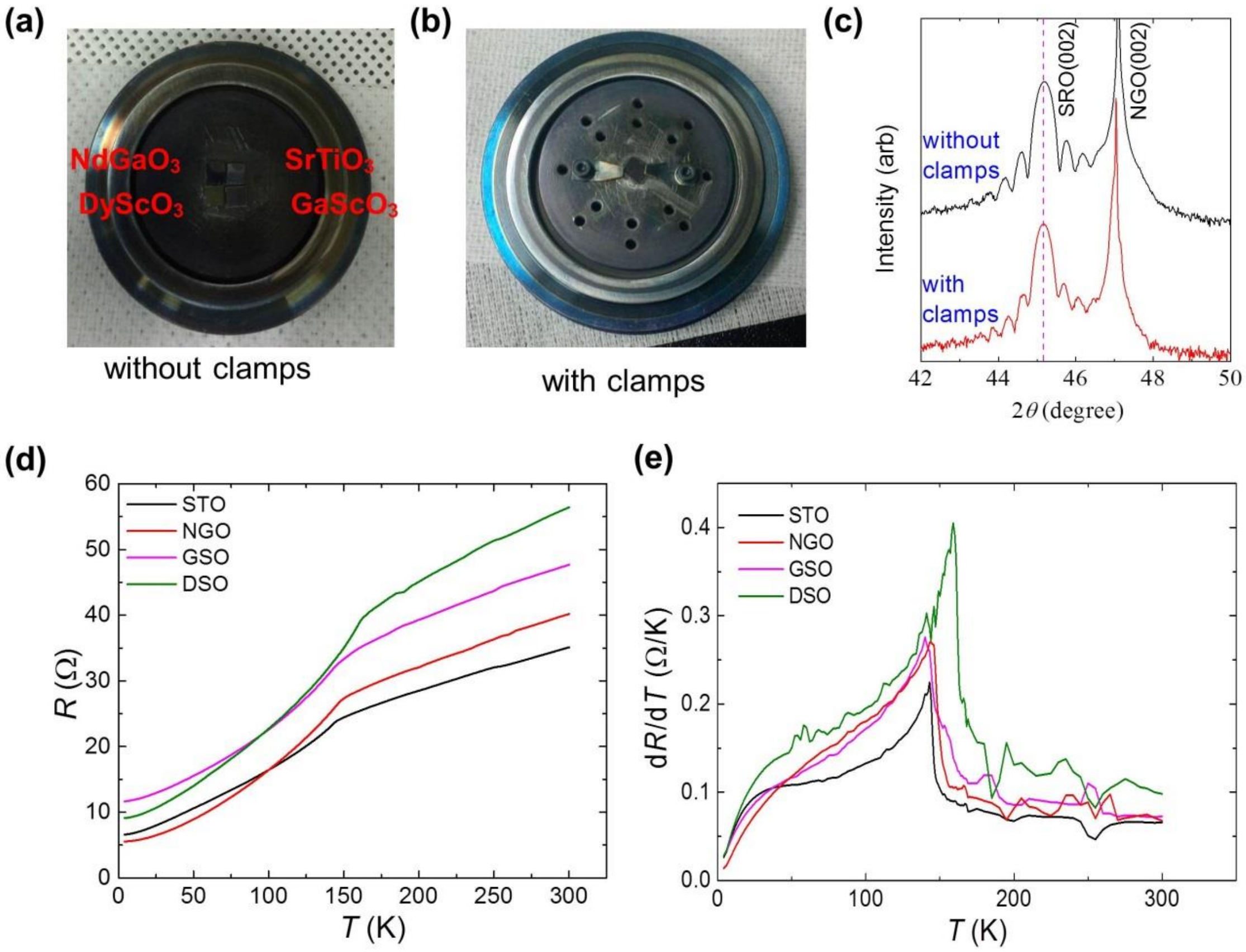

**Figure S3. Resistance vs temperature of SRO113 thin films deposited on various substrates.** (a) Temperature dependent resistance of 30-nm thick SRO113 thin films deposited on STO (black line), NGO (red line), GSO (magenta line), and DSO substrates (green line). RRR values of these films reveal the high quality of the films. (b) Temperature dependent derivative of resistance shows that the $T_{\text{Curie}}$ measure in $M(T)$ and $R(T)$ is closely correlated for these films.

**Table S1. Structural, electric and magnetic properties of 30-nm thick $SrRuO_3$ films deposited on various substrates.**

| SRO113 30-nm | Substrate *a*-axis (nm, pseudo-ubic) | In-plane strain (%) | RRR | $M_{r_c\text{-axis}}$ (emu/cm$^3$) @ 4 K | $T_{Curie}$ (K) from $R(T)$ | $T_{Curie}$ (K) from $M(T)$ |
|---|---|---|---|---|---|---|
| NGO Orthorhombic | 0.3864 | –1.68 | 7.3 | 214 | 146 | 148 |
| SRO214 (l4/mmm) | 0.3871 | –1.50 | | 458 | | 159 |
| STO Cubic | 0.3905 | –0.64 | 5.2 | 168 | 143 | 140 |
| DSO Orthorhombic | 0.3946 | 0.41 | 6.3 | 199 | 159 | 158 |
| GSO Orthorhombic | 0.397 | 1.02 | 4.1 | 153 | 142 | 142 |
| Nb:STO Cubic | | | | 100 | -- | 137 |

## 3. Reciprocal Space Mapping (RSM) of SRO113/SRO214 thin film

It was rather hard to obtain the X-ray RSM for SRO113/SRO214 because of somewhat irregular surface of SRO214 substrates on the opposite side of the film. Irregular surface makes it difficult to align the interface accurately along the desired in-plane direction. After a careful alignment we collected RSM for a 30-nm thick SRO113/SRO214 thin film. Figure S4 shows the peak from SRO214 at upper region and a weak peak signature for SRO113 at lower region.

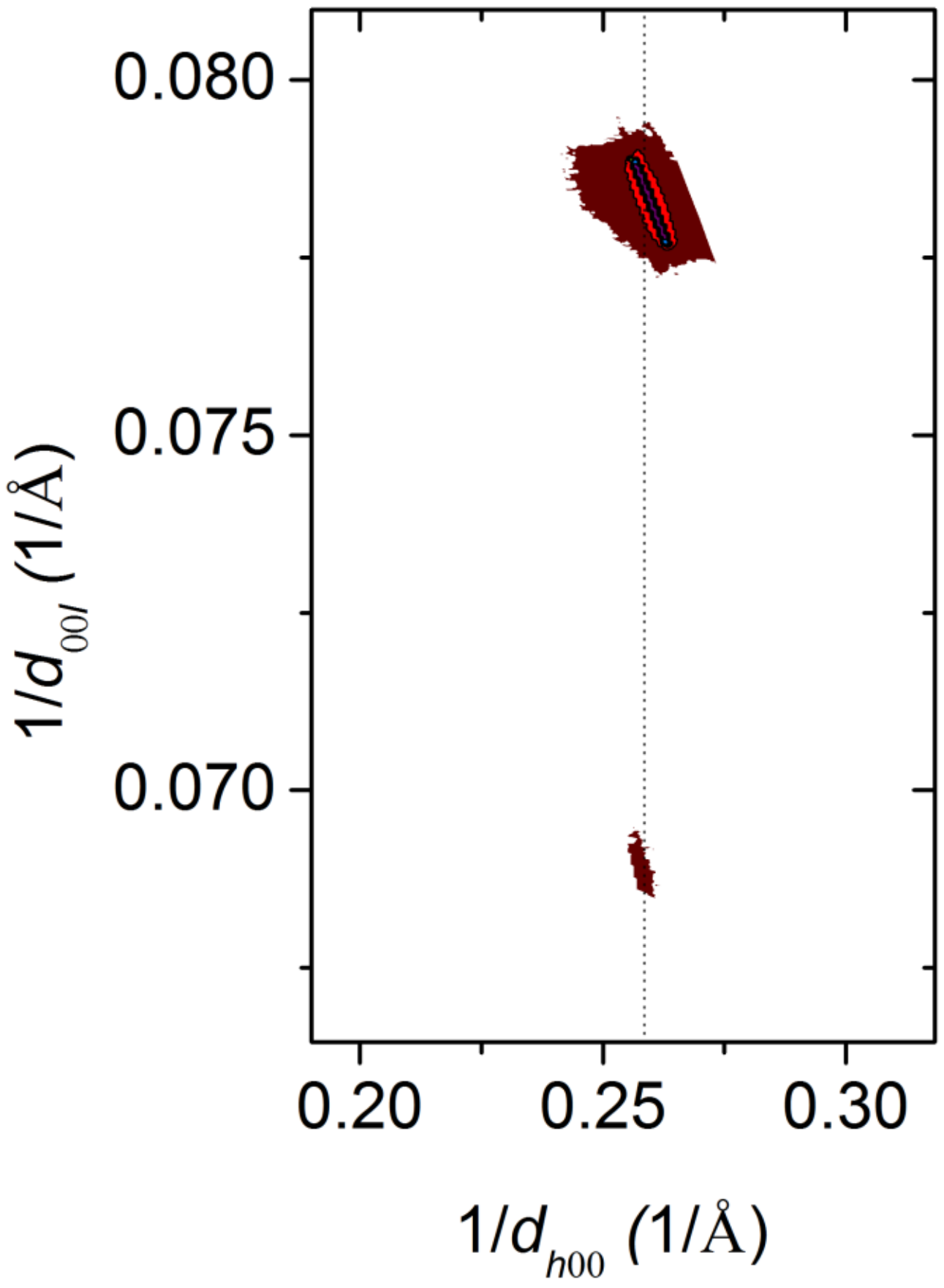

**Figure S4 RSM data for SRO113/SRO214 thin film.** X-ray reciprocal space mapping around (1 0 11) for a 30-nm thick SRO113/SRO214 system shows the peaks for SRO214 in the upper region and for SRO113 at lower region and.

## 4. Magnetization of SRO113/Nb:STO (5%)

It is demonstrated in the main text that enhanced magnetization is observed only when SRO113 thin films are deposited on a conducting substrate SRO214. In this case, non-trivial electronic correlation at the interface between metallic substrate and film may play a crucial role in enhancing the magnetization of SRO113/SRO214 thin films. To study such an effect, we used 5%-Nb doped $SrTiO_3$ (Nb:STO) metallic substrate. Note that Nb:STO substrate is much less conductive than SRO214 substrates. Interestingly, remnant magnetization and $T_{Curie}$ for a 30-nm SRO113 film deposited on a Nb:STO substrate are smaller than those for the SRO113/STO thin films, as compared in Figure S5. Thus a metallicity of the substrate itself does not enhance the magnetization; on the contrary for Nb:STO substrate it is suppressed.

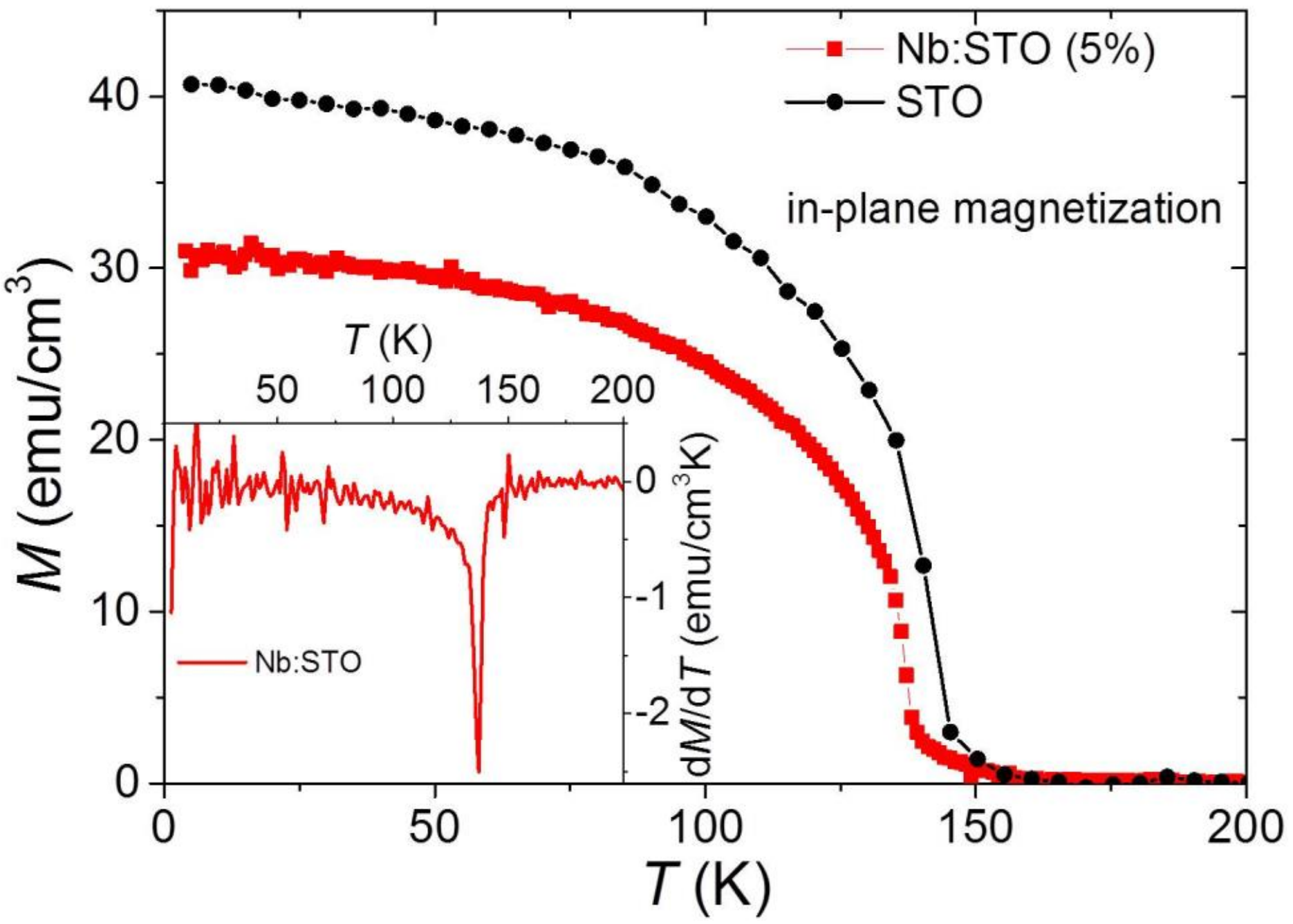

**Figure S5. Magnetization of SRO113 thin film deposited on Nb:STO and STO substrates.** Temperature dependence of remnant magnetization along the *ab*-plane (along the interface) of 30-nm SRO113 thin films deposited on Nb:STO substrate (red symbols) and STO substrate (black symbols). Remnant magnetization and $T_{Curie}$ are suppressed for SRO113/Nb:STO thin film compared to SRO113/STO film.